\documentclass{ws-procs9x6}

\begin{document}

\title{Detection of chirality and mutations of knots and links
}

\author{RAMADEVI PICHAI}

\address{Department of Physics, Indian Institute of Technology Bombay,\\
Mumbai,Maharashtra-400 076, INDIA\\
$^*$E-mail: ramadevi@phy.iitb.ac.in}

\begin{abstract}
In this brief presentation, we would like to present our attempts of
detecting chirality and mutations from Chern-Simons gauge theory.
The results show that the generalised knot invariants, obtained from
Chern-Simons gauge theory, are more powerful than Jones, HOMFLYPT and 
Kauffman polynomials. However the classification problem of knots and 
links is still an open challenging problem.
\end{abstract}


\bodymatter

\section{Introduction}\label{aba:sec1}
The classification of three and four manifolds is one
of the open questions which has been addressed by
both mathematicians and physicists. In particular,
physicists have shown that a class of quantum field theories
called topological field theories provides an elegant approach
to solve these problems.

\begin{figure}
\psfig{file=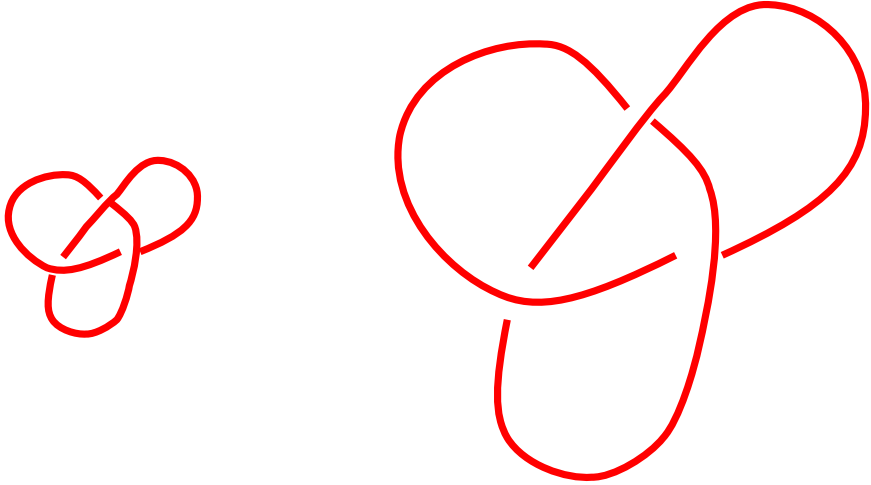,width=4.5in}
\caption{Both knots $C$ (though different sizes) are trefoil knots}
\label {aba:fig1}
\end{figure}

The main idea in any quantum field theory is to represent the
theory by an action $S$ which gives information about the
particle content and their interactions. The
interaction strengths are given by coupling constants.
For capturing the topological features of knots or links as
shown in Fig.~\ref{aba:fig1}, we need
a theory which does not change if we alter the
shape or size of these knots or links.
One such theory is the Chern-Simons gauge theory
where the action $S$ is explicitly metric independent.
Hence, Chern-Simons field theory provides a natural framework to study
knots,links and three manifolds.
The action $S$ defining the Chern-Simons theory on a three manifold
$M$ based on a gauge group $G$ is
\begin{eqnarray}
S&=&{k \over 4 \pi}\int_M Tr\left(A \wedge dA+{2 \over 3}A \wedge A \wedge A
\right)\nonumber\\
~&=&~
\frac{k}{4 \pi} \int_M \epsilon_{\mu \nu
\lambda}~ d^3 x ~Tr \left(A_{\mu} \partial_{\nu}
A_{\lambda} + \frac{2}{3} A_{\mu} A_{\nu}  A_{\lambda}\right)~,
\end{eqnarray}
where $k$ is the coupling constant and $A_{\mu}$'s are the
gauge fields or connections matrix-valued in group $G$.

The knots or links - for example, the trefoil knot $C$  carrying
representation $R$ of the gauge group $G$
are described by the expectation value
of Wilson loop operators $W_R(C)= Tr[Pexp\oint A_{\mu} dx^{\mu}]$:
\begin{equation}
V_R[C]=\langle W_R(C) \rangle~=~
\frac{ \int_M [{\cal D}A] ~ W_R(C)~ exp(i S)}{
{\cal Z}[M]}~,
\end{equation}
where $${\cal Z}[M]=\int_M [{\cal D} A]~ exp(iS)$$ is the partition
function and $V_R[C]$ are the  knot invariants.

\noindent
Witten's pioneering work \cite{wit1} established a three-dimensional
definition for knots and links. In particular, Jones and HOMFLYPT
polynomials and their recursion relations were obtained from
Chern-Simons gauge theory based on $SU(2)$ and $SU(N)$ gauge groups.
We can relate coupling constant $k$ and the
rank $N$ to the polynomial variables of
HOMFLYPT polynomials. Similarly, the Jones' polynomial variable
$q$ will be related as $q=\exp\left(2 \pi i / k+2\right)$.

\begin{figure}
\psfig{file=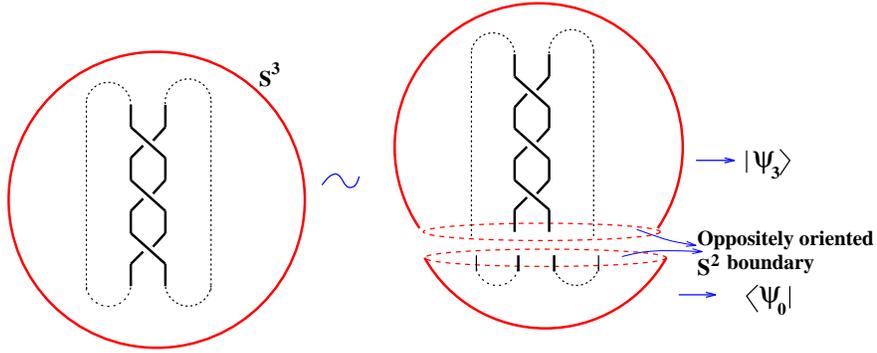,width=4.5in}
\caption{(a) Trefoil in $S^3$ drawn as a closure of braid $\equiv$ (b) gluing of two three-balls with oppositely oriented $S^2$ boundaries}
\label {abb:fig2}
\end{figure}

\noindent
The two main ingredients which go into the evaluation of the polynomial
invariants $V_R[C]$ of knots and links are
\begin{enumerate}
\item Connection between Chern-Simons theory on the three-dimensional
ball to the two-dimensional
Wess-Zumino conformal field theory
on the boundary of the three-ball.
\item Using Alexander's theorem, any knot or link
can be obtained as a closure of braid.
\end{enumerate}
In Fig.~\ref{abb:fig2}(a), we illustrate Alexander's theorem by
re-drawing the trefoil knot as a closure of two-strand braid with
three crossings. Also, we have diagrammatically shown in 
Fig.~\ref{abb:fig2}(b) that these
knots or links in $S^3$ can be viewed as gluing two three-balls
with oppositely oriented $S^2$ boundaries. In this
particular trefoil knot example, the $S^2$ boundary has four-punctures.
The connection between Chern-Simons theory and Wess-Zumino conformal
field theory states that the Chern-Simons functional integral
over a three-ball with a four-punctured $S^2$ boundary corresponds to
state $|\Psi_3 \rangle$ which represents four-point correlator
conformal block in the Wess-Zumino conformal field theory. The suffix
$3$ on the state is to indicate that the middle two-strands
are braided thrice. In fact, the punctures get exchanged whenever
the middle two-strands get braided. We can
denote the no-crossing four-punctured boundary state
as $\vert \Psi_0\rangle$ and apply a braiding operator ${\cal B}$
thrice to get the state $\vert \Psi_3\rangle$:
\begin{equation}
\vert \Psi_3\rangle= {\cal B}^3 \vert \Psi_0\rangle~.
\end{equation}
Similarly, the state for the oppositely oriented boundary
will be in the dual space. For the above example in Fig. 2(b),
the state is $\langle \Psi_0|$.
The knot invariant is
\begin{equation}
V_R[C]= \langle \Psi_0 |\Psi_3\rangle=
\langle \Psi_0 |{\cal B}^3|\Psi_0\rangle~. \label {knot1}
\end{equation}
In order to see the polynomial form, we need to expand the
state $|\Psi_0\rangle$ in an eigenbasis of the braiding operator ${\cal B}$.
In general for the four-punctured $S^2$ boundary, the braiding
can be either on the side two-strands or on the middle two strands.
For clarity, we will take the gauge group $G=SU(2)$.
\begin{figure}
\psfig{file=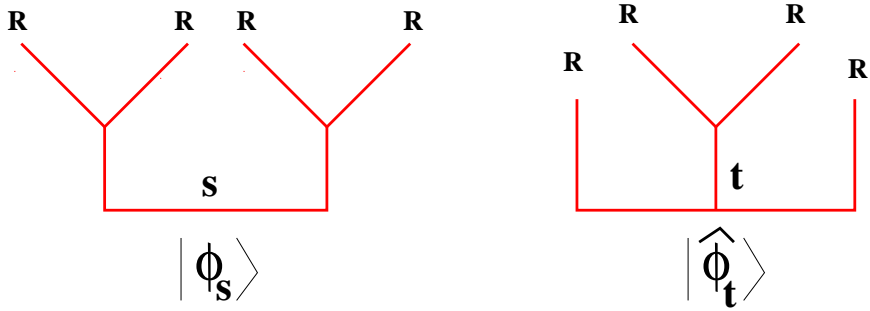,width=4.5in}
\caption{eigenbasis when braiding (a) side two-strands (b) middle two-strands.}
\label {abc:fig3}
\end{figure}
For the four-punctured $S^2$ boundary, we can chose eigenbasis
$|\phi_s\rangle$ if the braiding is in the side two-strands.
That is,
\begin{equation}
{\cal B}_1 |\phi_s\rangle = \lambda_s(R,R)|\phi_s\rangle ~,
\end{equation}
where the suffix $1$ on the braiding operator denotes the
braiding between first and the second strands and
the eigenvalue is $\lambda_s(R,R)$.
The basis is shown diagrammatically in Fig. 3(a) where $R$ denotes
the $SU(2)$ representation placed on the strands.
From the picture, the representation $s$ will be an element
in the tensor product $R\otimes R$. Similarly, for braiding
middle two strands, we choose the basis $|\hat{\phi}_t\rangle$ as
shown in Fig. 3(b) where $t \in R \otimes R$.
Clearly, these two basis states must be related by a duality
matrix:
\begin{equation}
|\hat{\phi}_t\rangle =a_{ts}\left[\begin{matrix}R&R\cr R&R
\end{matrix}\right] |\phi_s\rangle~.
\end{equation}
When the four strands carry the same representation, we can
write in shorthand notation the duality matrix as
\begin{equation}
a_{ts}\left[\begin{matrix}R&R\cr R&R
\end{matrix}\right] \equiv a_{ts}~.
\end{equation} 
These duality matrices turns out to be proportional to quantum Racah
coefficients. The explicit form of $SU(2)_q$ Racah coefficients
and their identities satisfied are available in 
Ref.~\refcite{reshi}. See also papers\cite{trg1,trg2,kaul}~. 
\noindent
In this example, the braiding involves middle 
two strands. Hence the state $\vert \Psi_0\rangle$ can be expanded
in the middle-strand basis $|\hat \Phi_t \rangle$:
\begin{equation}
\vert \Psi_0 \rangle ~=~\sum_t \mu_t~ \vert \hat \Phi_t \rangle~.
\end{equation}
\noindent
Interestingly, the coefficients  $\mu_t$ has to satisfy
\begin{equation}
\mu_t ~=~\sqrt{V_t[U]}~=~\sqrt{S_{0t}/S_{00}} \equiv \sqrt{dim_q t}~, 
\end{equation}
so that two equivalent knots share the same polynomial invariant.
Here, $V_t[U]$ denotes the polynomial invariant for unknot carrying
representation $t$ whose form can be written as the ratio of elements
of the modular transformation matrix $S$ in Wess-Zumino conformal field theory
or in terms of quantum dimensions of the representation $t$ of
the quantum group as indicated in the above equation. The knot invariant
(\ref{knot1}) will be
\begin{equation}
V_R[C]~=~ \langle \Psi_0 \vert {\cal B}^3 \vert \Psi_0 \rangle~=~
\sum_t dim_q t (\lambda_t (R,R))^3 ~.
\end{equation}
So far, we have not introduced orientation on the strands. In general, the
braiding eigenvalue depends on the {\bf framing} and also on the relative 
orientation on the two braiding strands. Two conventional 
framing are standard framing and blackboard framing. Standard
framing is one where the self-linking number of the knot with its
frame is zero. This is useful to obtain ambient isotopy invariants.
The self-linking number matches the crossing number in the blackboard
framing. Hence the braiding eigenvalue in the blackboard framing
is useful to obtain regular isotopy invariants. 
In the trefoil example, we could place parallel orientation in the 
middle two-strands. As the crossing sign due to braiding is positive, 
we called such a braiding as right-handed braiding. 
Similarly, an inverse braiding leading to mirror of trefoil ($T^*$)
will be called left-handed braiding. 

\begin{figure}
\psfig{file=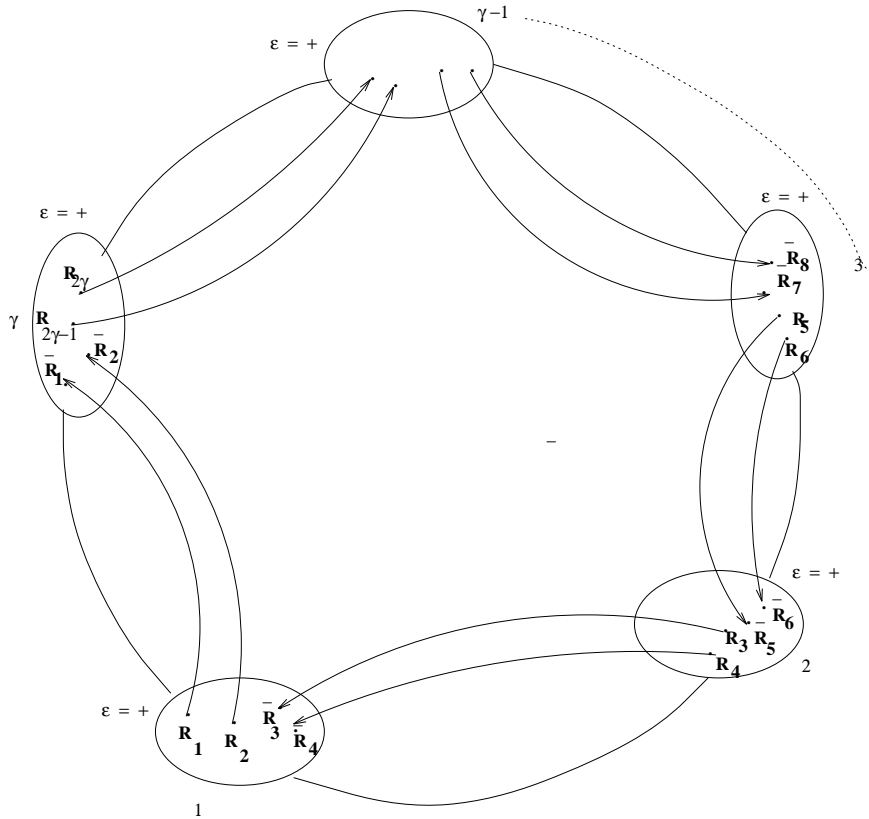,width=4.5in}
\caption{Three-ball with $r$ $S^2$ boundaries each with 
four-punctures.}
\label {abd:fig4}
\end{figure}

\begin{figure}
\psfig{file=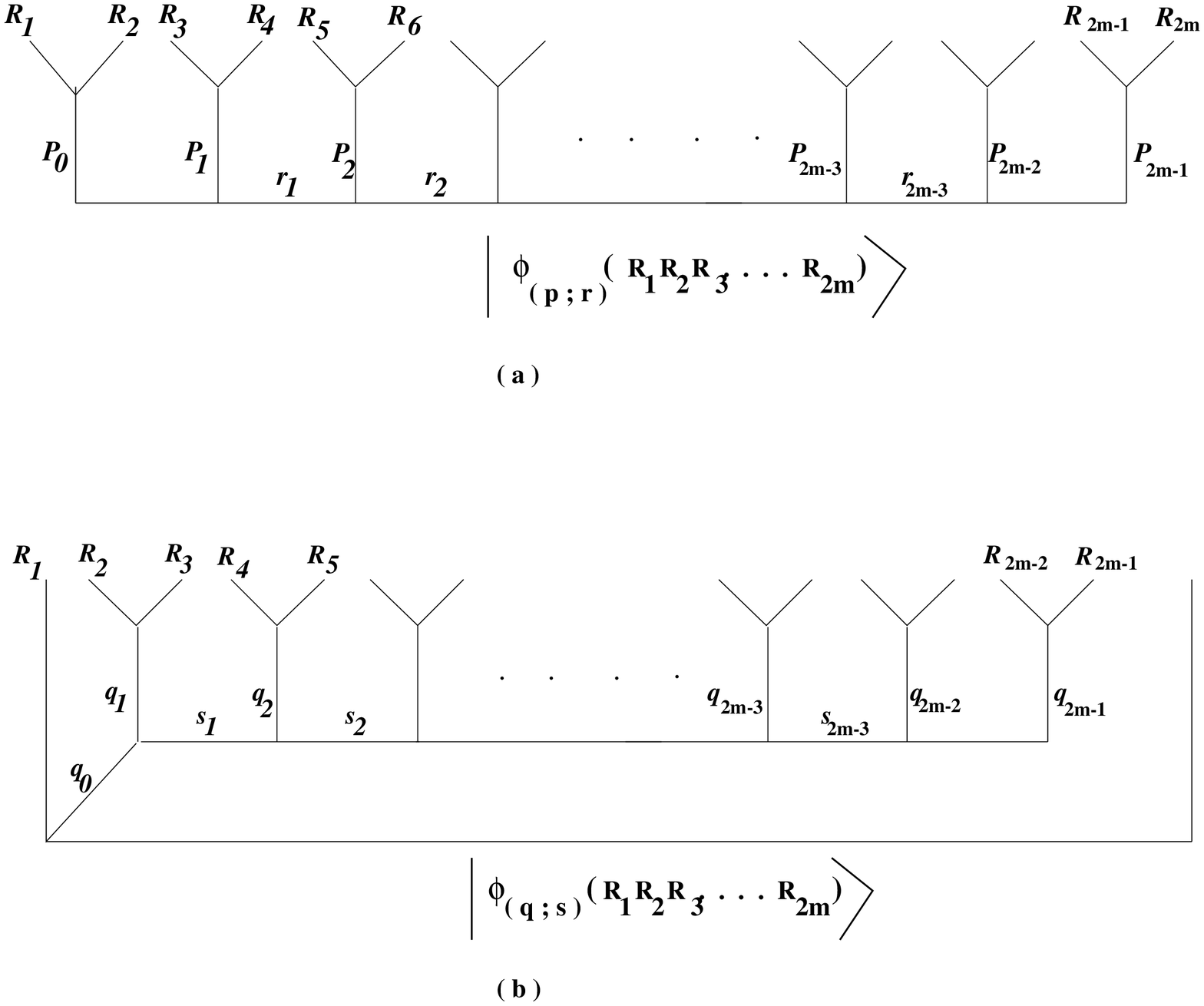,width=4.5in}
\caption{Basis states for a $n$-punctured $S^2$ boundary.} 
\label {abe:fig5}
\end{figure}

For parallely oriented strands, the right-handed braiding eigenvalue 
in standard framing is
\begin{equation}
\lambda_t^{(+)}(R,R)~=~(-1)^{\epsilon}
q^{2 C_R - C_t/2},~  q= e^{2 \pi i \over k+C_v}.
\end{equation}
where $C_R,C_t,C_v$ denotes the quadratic casimirs in the
respective $R,t$ and adjoint representation. $\epsilon$ will 
be $\pm 1$ depending on the representation $t$ appears 
symmetrically or antisymmetrically in the tensor product
$R \otimes R$. Similarly, the left-handed braiding eigenvalue
for antiparallely oriented strands is
\begin{equation}
\lambda_t^{(-)}(R,R)~=~(-1)^{\epsilon}
q^{C_t/2},~  q= e^{2 \pi i \over k+C_v}.
\end{equation}
Now using the appropriate braiding eigenvalues, the 
knot invariants can be written as polynomials in the
variable $q$. The method for a four-punctured $S^2$ boundary is 
generalisable for $r$  such four-punctured $S^2$ boundaries as
shown in Fig.~\ref{abd:fig4}.
We will see in the next section that this building block will be 
useful to redraw knots like knot $9_{42}$ and knot $10_{71}$ as gluing of three-balls with one or more four-punctured $S^2$ boundaries.
The basis state for such a $r$-$S^2$ boundaries is 
\begin{equation}
\nu_r \ = \ \sum_{R_s} \frac{\vert {\phi_{R_s}^{(1)side} }\rangle 
\vert {\phi^{(2)side}_{R_s} }\rangle \cdots 
\vert {\phi_{R_s}^{(r)side} }\rangle} {(dim_q R_s)^{\frac{r-2}{2}}}~,
\label {rbdy}
\end{equation}
where $R_s \in R_i \otimes R_{i+1}$ for any $i$. Sometimes, it is useful to 
keep $S^2$ boundaries with more than four-puctures. Then the 
basis state for a $S^2$ boundary with $n$ puctures
will be a $n$-point conformal block. For braiding ${\cal B}_{2i+1}$'s
and ${\cal B}_{2i}$'s, we choose the basis  
shown in Fig.~\ref{abe:fig5}(a) and (b) respectively.
The procedure we elaborated for the trefoil obtained
from gluing two three-balls each with a four-punctured
$S^2$ boundary is generalisable for any knot.
That is, using the building blocks in Fig.~\ref{abd:fig4} 
or three-balls with $n$-punctured $S^2$ boundaries,
it is not difficult to see that any knot can be 
obtained either from gluing two three-balls each with a $n$-punctured
$S^2$ boundaries  or from gluing many three-balls with one or
more $r$ four-puctured $S^2$ boundaries.
The method presented here enables direct evaluation of 
any knot/link polynomial directly without going through the 
recursive procedure.

Though we have concentrated on the gauge group $SU(2)$, it is
straightforward to generalise for any compact semi-simple gauge 
group\cite{rama}. The representation $R$ must be replaced 
by conjugate representation $\bar R$ ($R \equiv \bar R$ for $SU(2)$) 
depending on the oriented strand is outgoing from 
or incoming to a $S^2$ boundary as shown 
in Fig.~\ref{abd:fig4}.
Thus, we can place any representation $R$ of any compact 
semi-simple gauge group on the knot and obtain 
\underline{\bf generalised knot invariants.} If we place
the defining representation on the strands, we recover
some of the well-known polynomials as tabulated below: 

\noindent
\begin{tabular}{|l|l|}\hline
Gauge Group&Polynomial\\\hline 
$SU(2)$&Jones'\\
$SU(N)$&Two-variable HOMFLYPT\\
$SO(N)$&Two-variable Kauffman\\
\hline
\end{tabular}

\begin{figure}
\psfig{file=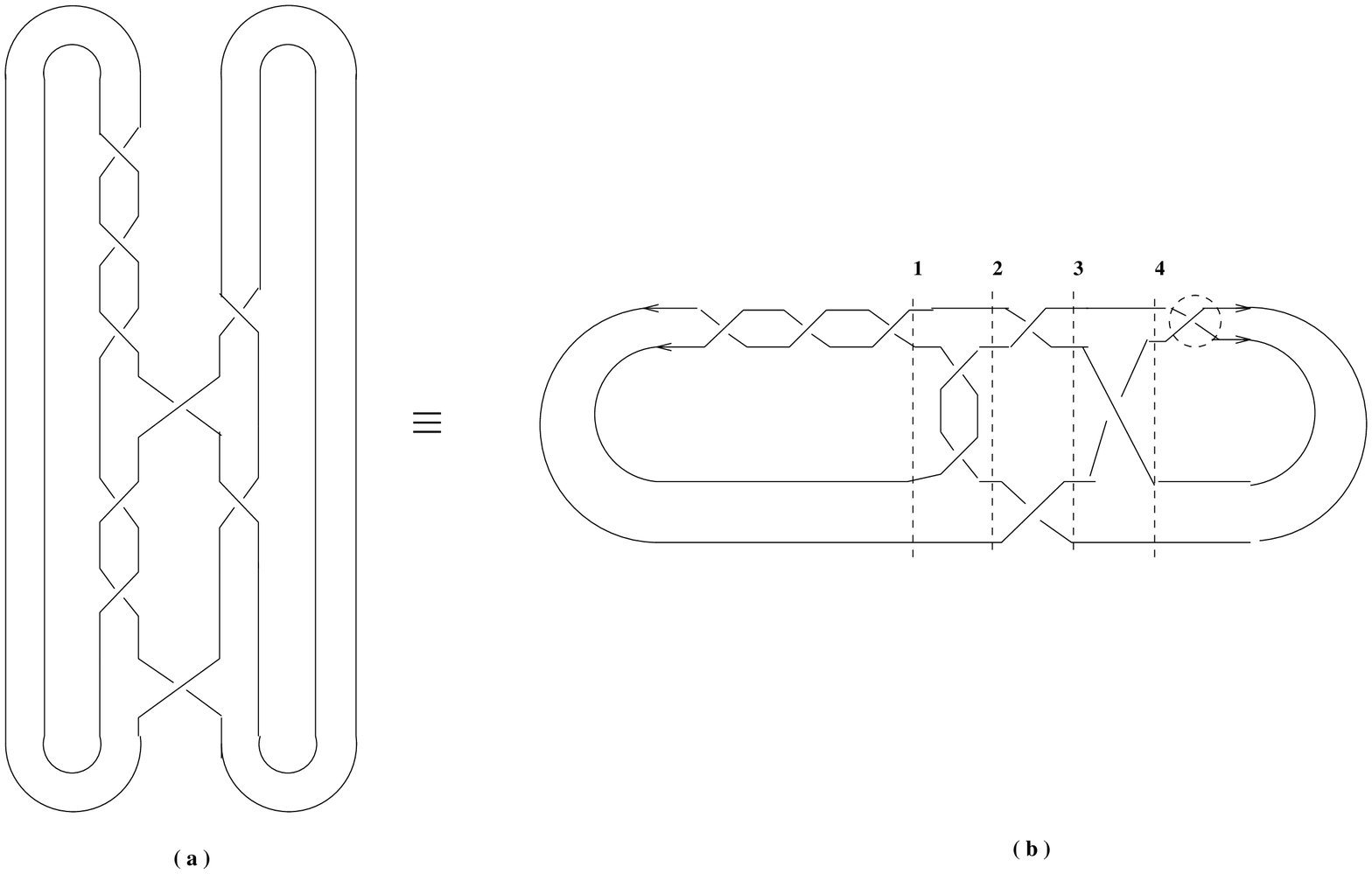,width=4.75in}
\caption{Chiral Knot $9_{42}$} 
\label {abf:fig6}
\end{figure}

\noindent
We know that these well-known polynomials do not solve the 
classification problem. Apart from these special cases,
Chern-Simons field theory gives a huge pool of generalised
polynomials depending on placing representation $R$ of any 
gauge group other than the defining representation on 
the strands. We believe that at least one of these generalised
Chern-Simons invariant will be able to distinguish two inequivalent
knots which are not distinguished by the well-known polynomials.

Knot theory literature gives a list of chiral knots and mutant
knots which are not distinguished by Jones', HOMFLYPT and Kauffman. 
We tried to check the ability of generalised knot invariants, 
from Chern-Simons field theory, to detect {\bf chirality and mutations}.
We address the chirality detection in the following section.
\section{Chirality Detection}
\begin{figure}
\psfig{file=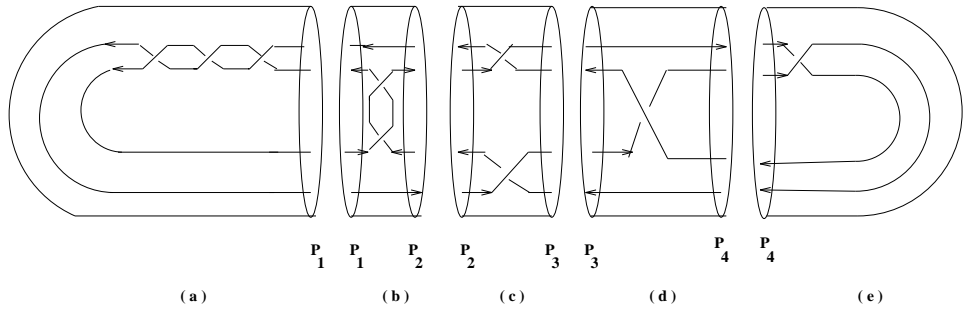,width=5in}
\caption{Gluing of five building blocks} 
\label {abg:fig7}
\end{figure}

Upto 10 crossings, there are two knots : knot $9_{42}$ 
and knot $10_{71}$ which are chiral but their
chirality is not detected by the well-known polynomials.
In Fig.~\ref{abf:fig6}, we have drawn {\bf knot $9_{42}$} in 
two equivalent ways.
Clearly, the knot $9_{42}$ can be obtained
as gluing of five building blocks as shown in Fig.~\ref{abg:fig7}.

For $G=SU(2)$,
\setlength{\unitlength}{1cm}
$R_n$ = 
\begin{picture}(2,.7)
\put(0,0){\line(1,0){1.8}}
\put(0,.3){\line(1,0){1.8}}
\multiput(0,0)(.3,0){7}{\line(0,1){.3}}
\put(.7,.5){\vector(-1,0){.7}}
\put(1.1,.5){\vector(1,0){.7}}
\put(.8,.5){$n$}
\end{picture} (spin $n/2$ representation) placed on knot, the states
for the building blocks can be written down following the 
methods presented in the previous section and also using the properties
of the duality matrix\cite {trg1,trg2,kaul}.
The states for these five building blocks are \cite{rama1}
\begin{eqnarray*}
\nu _1(P_1)&=&\sum _{l_1=0}^n\sqrt
{[2l+1]}(-1)^{3(n-l_1)}q^{-3/2[n(n+2)-l_1(l_1+1)]}\vert {\phi_{l_1}^{(1)}}\rangle\\
\nu _1(P_4)&=&\sum _{l_5=0}^n(-1)^{n-l_5}q^{-1/2[n(n+2)-l_5(l_5+1)]}\vert
{\phi _{l_5}^{(1)}}\rangle\\
\nu _2(P_1;P_2)&=&\sum _{i_1,j_1,l_2,r=0}^n {a_{l_1r}a_{j_1r}a_{l_2r}\sqrt
{[2l_2+1]}\over \sqrt {[2r+1]}} \times 
q^{n(n+2)-l_2(l_2+1)}\vert {\phi _{i_1}^{(1)}}\rangle \vert{\phi _{j_1}^{(2)}}
\rangle\\
\nu _2(P_2;P_3)&=&\sum _{l_3=0}^nq^{l_3(l_3+1)}\vert {\phi _{l_3}^{(1)}}
\rangle \vert {\phi _{l_3}^{(2)}}\rangle\\
\nu _2(P_3;P_4)&=&\sum
_{i_2,j_2,l_4=0}^n(-1)^{l_4}q^{-l_4(l_4+1)/2}a_{l_4i_2}a_{l_4j_2}\vert
{\phi _{i_2}^{(1)}}\rangle \vert{\phi _{j_2}^{(2)}}\rangle~.
\end{eqnarray*}
Here $P_i$'s denote the $S^2$ boundaries as indicated in Fig.~{abg:fig7}.
Using the above states, the knot invariant is
\begin{eqnarray*}
V_n[9_{42}]& = &(-1)^n q^{{-3\over 2}[n(n+2)]}\sum
_{r,l_1,l_2,j_1,j_2=0}\sqrt {[2l_1+1]} \times\\
~&~&\sqrt {[2l_2+1]}\sqrt
{[2j_2+1]}
a_{l_1r}a_{l_2r} a_{j_1r}a_{j_1j_2} \times \\
~&~&(-1)^{l_1}q^{{3\over 2}[l_1(l_1+1)]}
q^{{3\over 2} [j_1(j_1+1)]}q^{-l_2(l_2+1)}q^{j_2(j_2+1)}
\end{eqnarray*}
\begin{figure}
\psfig{file=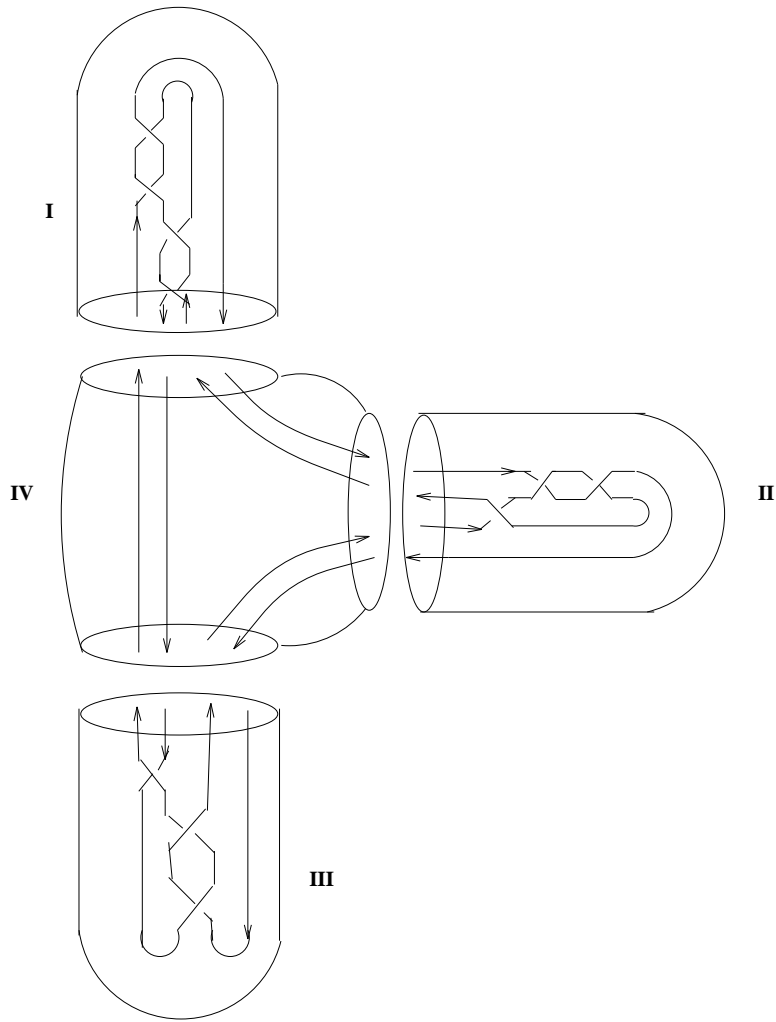,width=4in}
\caption{Chiral Knot $10_{71}$ obtained from gluing the building blocks} 
\label {abh:fig8}
\end{figure}
We checked the general result for the special cases. That is,
$n=1$ gives Jones' polynomial and $n=2$ gives Akutsu-Wadati/Kauffman 
polynomial\cite{akut}. Interestingly, for $n=3$, the polynomial is 
\begin{eqnarray*}
V_3[9_{42}]&=&
q^{45/2}-q^{41/2}-q^{39/2}+q^{35/2}+q^{23/2}+q^{21/2}-q^{19/2}\\
~&~&-
q^{17/2}+q^{13/2}-q^{9/2}+q^{5/2}+q^{3/2}+ q^{-3/2}+ q^{-5/2}\\
~&~&-q^{-13/2}-q^{-15/2}+q^{-21/2}+2q^{-23/2}-q^{-27/2}+2q^{-31/2}\\
~&~&-3q^{-35/2}-q^{-37/2}+q^{-39/2}+q^{-41/2}~.
\end{eqnarray*}
Clearly,  $V_3[9_{42}] (q) \neq V_3[9_{42}] (q^{-1})$ indicating that
$SU(2)$ Chern-Simons spin $3/2$ ($n=3$ in representation $R_n$) 
knot polynomial is powerful to detect chirality.
Similar exercise was performed for knot $10_{71}$ by gluing the
four building blocks as shown in
Fig.~\ref{abh:fig8}. The knot invariant \cite {rama1} is
\begin{eqnarray*}
V_n[10_{71}]&=&
(-1)^n q^{{n(n+2)\over 2}}\sum _{i,r,s,u,m=0} \sqrt
{{[2r+1][2s+1][2u+1]\over [2m+1]}}a_{im}\\
~&~&a_{ms}a_{rm}a_{iu}(-1)^s~q^{-i(i+1)}q^{m(m+1)}q^{-r(r+1)}q^{u(u+1)}
q^{{3\over 2}s(s+1)}~.
\end{eqnarray*}
For $n=3$, we have checked that $V_3[10_{71}](q) \neq V_3[10_{71}](q^{-1})$ 
confirming the ability of generalised Chern-Simons invariant in 
detecting chirality. With such positive results
for chiral knots, we attempted to check whether generalised
Chern-Simons invariant is capable of detecting mutation operation
which we shall present in the following section.
\section{Mutation and mutant knots}
\begin{figure}
\psfig{file=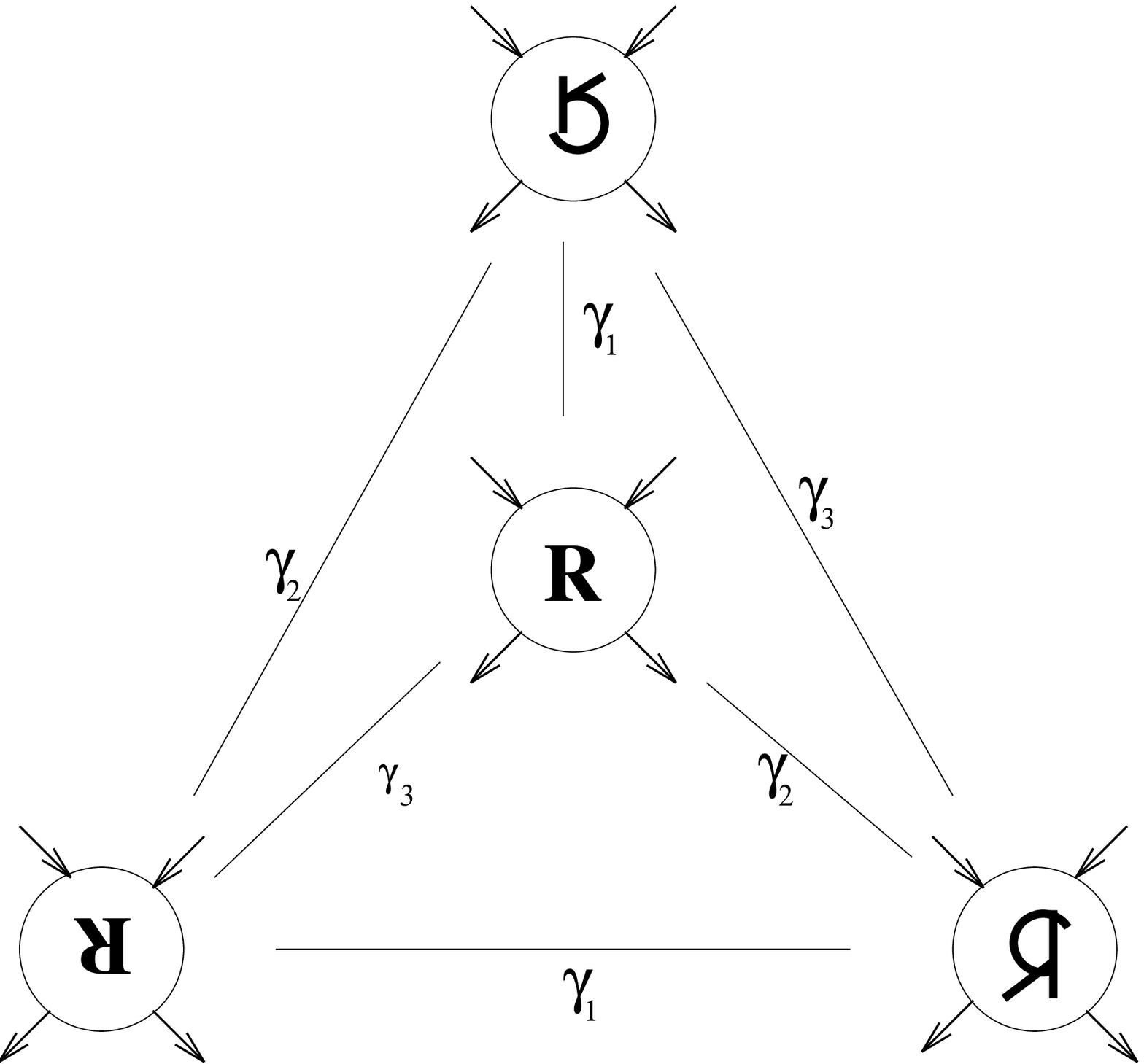,width=3in}
\caption{Mutation Operation} 
\label {abi:fig9}
\end{figure}

\begin{figure}
\psfig{file=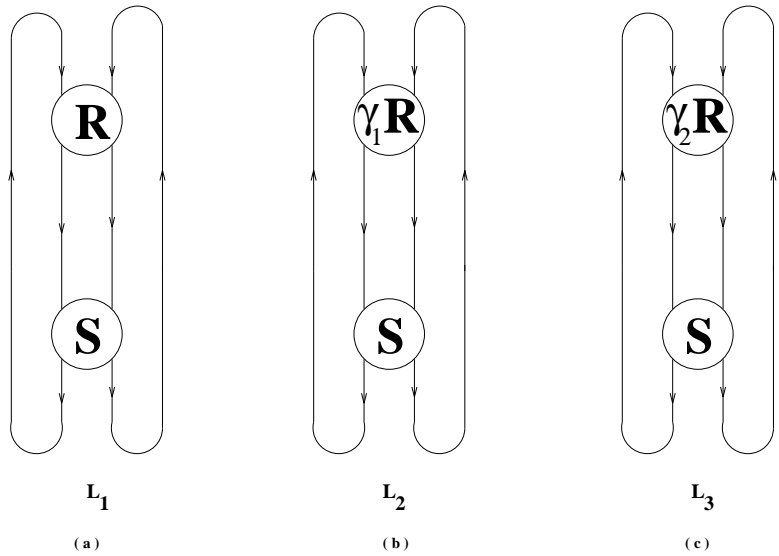,width=4in}
\caption{Mutant Knots} 
\label {abj:fig10}
\end{figure}

\begin{figure}
\psfig{file=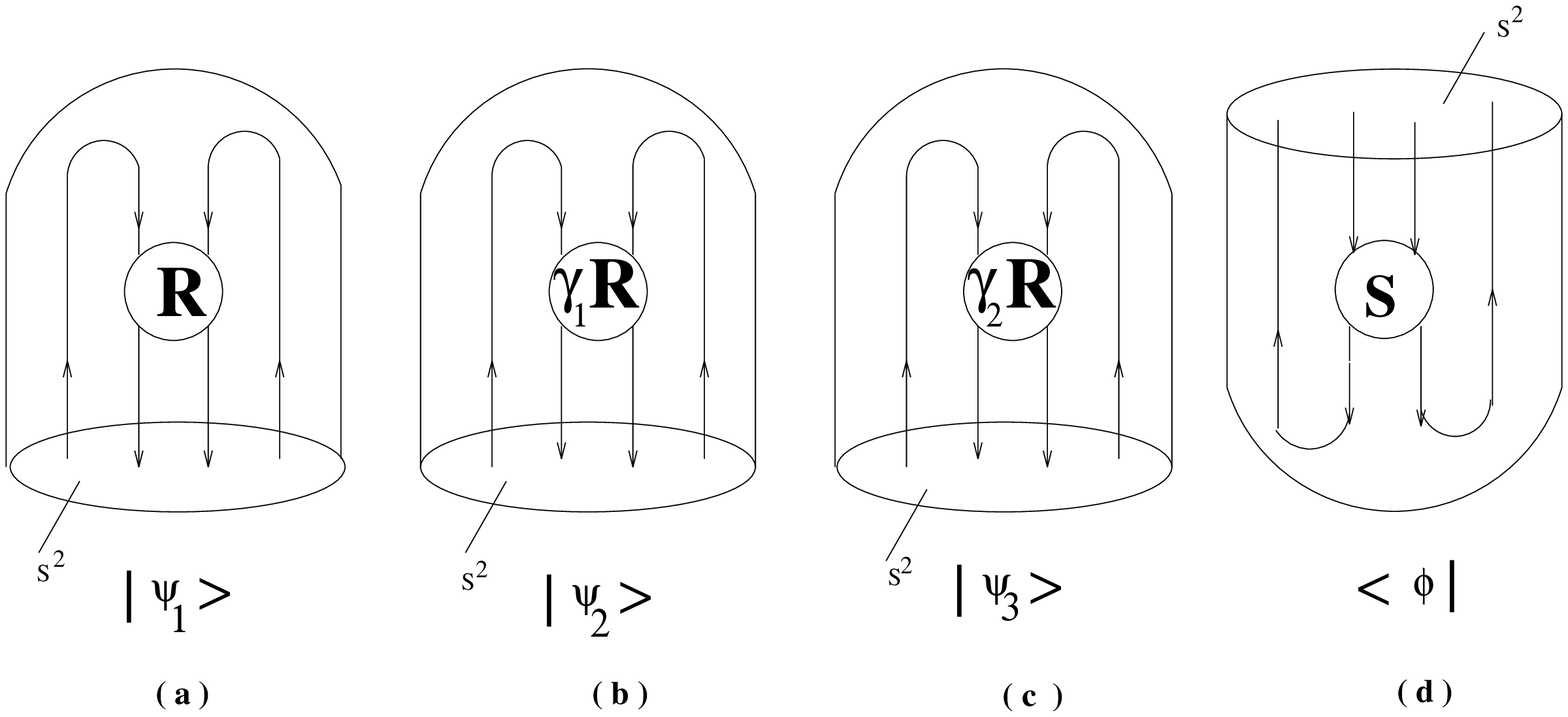,width=5in}
\caption{Building blocks} 
\label {abk:fig11}
\end{figure}

\begin{figure}
\psfig{file=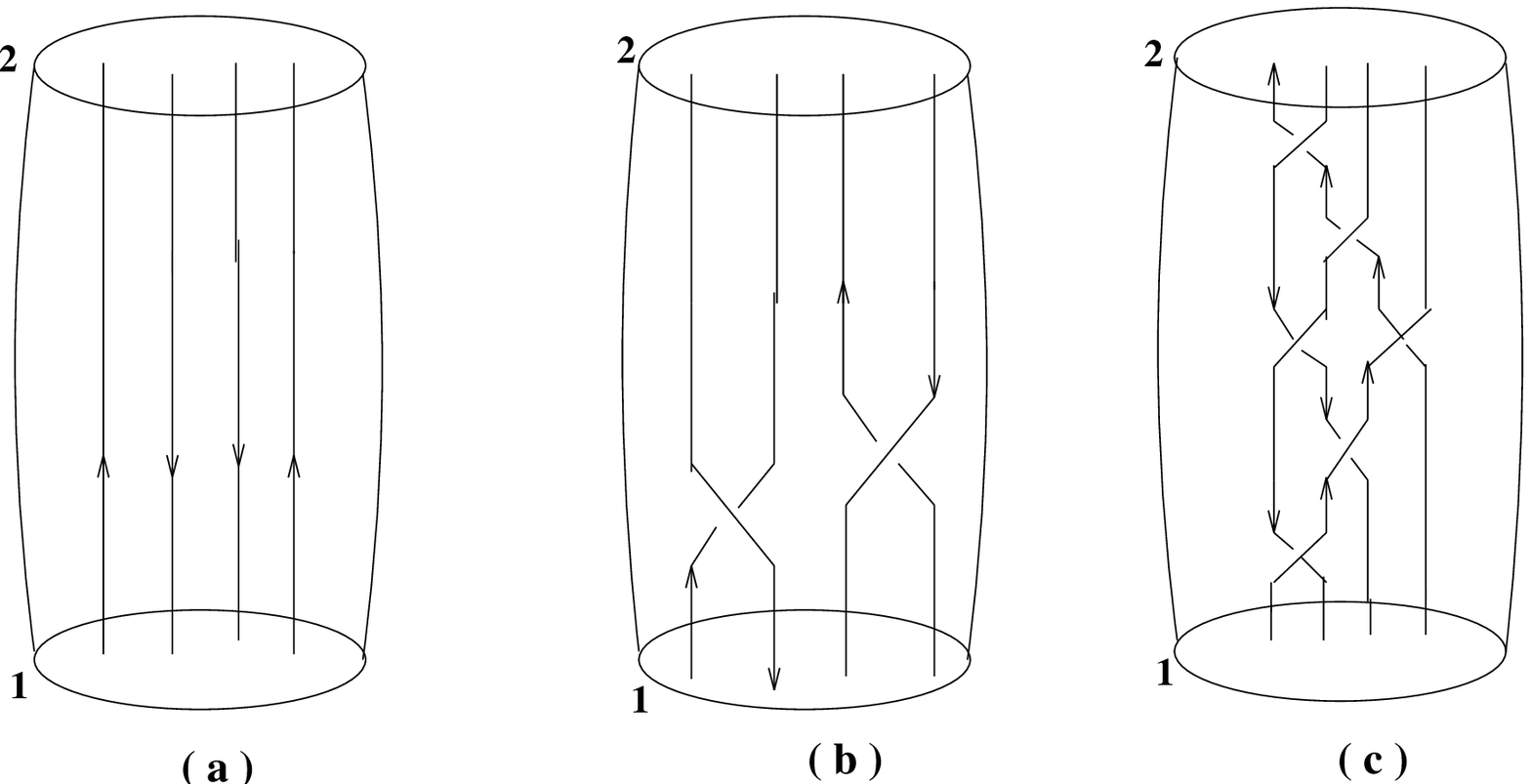,width=5in}
\caption{ state (a)  $\nu_1$ (b) $\nu_2$ and 
(c) $\nu_3$~.}
\label {abl:fig12}
\end{figure}

Remove a two-tangle region from any knot and do 
a rotation by $\pi$ about any of the three perpendicular
axis. Replace the rotated two-tangle back with
suitable reversal of orientation of the strands to give another
knot. This operation is called mutation and the 
two knots are said to be mutants. In 
Fig.~\ref{abi:fig9}, we have indicated the three mutation operation 
$\gamma_i$'s.
An example for the mutant knots is the well-known
eleven-crossing Kinoshita-Terasaka and Conway knots.
We can formally represent mutant knots shown in Fig.~\ref{abj:fig10} 
as gluing two three-balls with the two-tangle rooms as shown in 
Fig.~\ref{abk:fig11}.
That is, gluing Fig.~\ref{abk:fig11}(d) with any of the
Fig.~\ref{abk:fig11}(a),(b) 
and (c) gives mutant knots.
Interestingly, the states (a), (b), (c) in Fig.~\ref{abk:fig11} can be obtained
by gluing Fig.~\ref{abk:fig11}(a)
with the two-boundary states (a), (b) and (c) in Fig.~\ref{abl:fig12}
respectively.
Clearly, $\nu_2$ and $\nu_3$ represent mutation $\gamma_1$
and $\gamma_2$ respectively in Fig.~\ref{abi:fig9} respectively.
As braid words, Fig.~\ref{abl:fig12} (a), (b), (c) are different but the
states $\nu_i$'s are same:
\begin{equation}
\nu_2~=~
\sum _l~\vert {\phi_l^{side(1)}}\rangle b_1 b_3^{-1} \vert
{\phi_l^{side(2)}}\rangle~=~{\cal C}\nu_1~. 
\end{equation}
where we have used
$$b_1\vert  {\phi_l^{side}}\rangle = b_3 \vert {\phi_l^{side}}\rangle=
\lambda_l^{(-)}(R,\bar R) \vert {\phi_l^{side}}\rangle~,$$
and the operator ${\cal C}$ interchanges the representations
on the first and second, the third and fourth punctures in that basis.
Similarly, we can show
\begin{equation}
\nu_3~=~\sum_l \vert {\phi_l ^{side (1)}}\rangle b_1b_2b_1b_3b_2b_1
  \vert {\phi_l^{side (2)}}\rangle= {\cal C} \nu_1~.
\end{equation}
So, the {\bf generalised invariants} of the
mutants $L_1,L_2,L_3$ obtained by gluing Fig.~\ref{abk:fig11}(a),(b) and 
(c) with (d) are {\bf same}. 

It appears that the identities of the states for {\bf four-punctured 
boundaries} play a crucial role in making the states representing
the mutation operation $(\nu_2,\nu_3)$ to be similar to identity
braid $\nu_1$. In order to go beyond four-punctured boundary state,
we studied composite braiding \cite{rama2}~. 
Using the representation theory of composite
braids, we showed that the composite invariant for knots are
sum of the generalised knot invariants. This implies
composite invariants cannot detect mutations in knots.
However, some mutant links can be 
distinguished by composite invariants.

\section{Summary and Discussion}
In this article, we have briefly presented the direct evaluation
of generalised invariants of knots and links from Chern-Simons field
theory. For $SU(2)$ gauge group in Chern-Simons theory, we get the 
the colored Jones' polynomials for the knots and links carrying higher
dimensional $SU(2)$ representations. There is a huge pool of 
generalised Chern-Simons invariants for other gauge groups
like $SU(N)$, $SO(N)$ etc with knots and links carrying
arbitrary representations. The hope is that at least
one of the invariants will be able to distinguish two inequivalent
knots. 

We showed that the chiral knots upto 10 crossings are distinguished by the 
Chern-Simons field theory invariants. However, we gave a proof that 
the process of mutation cannot be detected within the Chern-Simons 
field theoretic framework. The proof is for knots and links
carrying arbitrary representation of any compact semi-simple
gauge group.

Discussions during the knot theory conference revealed
that there are approaches from quantum groups\cite{morton} and 
Floer homology\cite{gillam} whose invariants does distinguish 
11-crossing Kinoshita-Terasaka and Conway mutant knot.  
It will be interesting to to check whether this method distinguishes other
mutant knots.

\end{document}